\documentclass[aps,twocolumn,groupedaddress,showpacs,showkeys]{revtex4-1}
\usepackage{amsmath}
\usepackage[colorlinks, citecolor=blue,urlcolor=blue, linkcolor=red]{hyperref}
\hypersetup{linkcolor=blue}
\usepackage{amsmath}
\usepackage{amsfonts}
\usepackage{graphicx,color}
\usepackage{color}
\usepackage[bottom]{footmisc}
\usepackage{epstopdf}
\usepackage{epsfig}
\usepackage{graphics}
\usepackage[caption=false]{subfig}
\usepackage{url}
\usepackage{placeins}
\makeatletter

\newcommand{\Rmnum}[1]{\expandafter\@slowromancap\romannumeral #1@}
\makeatother

\begin{document}
\title{Single phonon state of mechanical mode via photon subtraction}
\author{M. Miskeen Khan}
\email[Electronic Address:\,]{mmiskeenk16@hotmail.com}
\author{M. Javed Akram}
\email[Electronic Address:\,]{mjakram@qau.edu.pk}
\author{F. Saif}
\email[Electronic Address:\,]{farhan.saif@fulbrightmail.org}
\affiliation{Department of Electronics, Quaid-i-Azam University, 45320 Islamabad, Pakistan}
\date{\today}

\begin{abstract}
We prepare single phonon Fock state of mechanical mode in optomechanical system based on photon subtraction process in the linear regime of optomechanical interaction. The correlation set between cavity and field modes is utilized to recast the state of a pre-cooled mechanical mode into a single phonon Fock state as the field mode is subjected to a photon subtraction type measurement at a dynamical instant of interaction time. The resultant conditional state of mechanical mode quantified by its Wigner function, exhibits an excellent compatibility with the standard single quantum Fock state. We confirm the nature of obtained state by  calculating the phonon statistics, and conditional fidelity which approaches to unity for an instantaneous measurement time within an appropriate choice of parametric regime. Moreover, we also study the effects of temperature parameter for the preparation of target state. The present scheme can be realized to engineer the  single quantum Fock state of macroscopic mechanical mode with currently available technology.  
\end{abstract}
\maketitle

\section{Introduction\label{I}}
Bizarre, often counter-intuitive, manifestations of quantum mechanics are generally observed at microscopic level. Recent rapid developments in quantum optomechanics \cite{aspelmeyer2014cavity} show that it is equally possible to observe these effects even at  macroscopic scale. Following a full set of quantum mechanical laws, the macroscopic optomechanical systems act as efficient workhorse for quantum control at large scale. Endeavors have been put for the generation of fully entangled states in simple \cite{vitali2007optomechanical} and hybrid optomechanical systems \cite{Asjad20122608,rogers2014hybrid,Kurizki31032015} consisting on field, mechanical and collective atomic modes. In addition, electromagnetically induced transparency \cite{PhysRevA.81.041803,0953-4075-48-6-065502,PhysRevA.90.023817}, slow and fast light effects \cite{safavi2011electromagnetically,PhysRevA.92.023846, akram2015efficient} have been reported in these hybrid systems as well. Hence these systems provide promising setups for quantum information processing \cite{stannigel2012optomechanical}. Apart from this, progressive theoretical  \cite{schliesser2006radiation, marquardt2007quantum,PhysRevA.77.011804,PhysRevA.90.033838,PhysRevLett.99.093902} as well as experimental \cite{chan2011laser,teufel2011sideband} developments have been made for putting mechanical oscillator into quantum ground state, which is feasible for making high-precision measurements with mechanical probes \cite{aspelmeyer2013cavity}. 

Non-classical states of macroscopic mechanical oscillator, on the other hand, have important practical advantages. For instance, a subclass of non-classical states is Gaussian mechanical squeezed states \cite{PhysRevA.88.023813,huang2009squeezing,asjad2014robust}, i.e. states which have Gaussian Wigner function with uncertainty below the zero point level into one of its motional quadrature. Such mechanical squeezed states are quite useful for ultra low force detection \cite{caves1980measurement}. The other subclass of non-classical states have non-Gaussian Wigner function, hence termed as non-Gaussian states. Such states generally have negative valued Wigner function and open a possibility to investigate the paradigm of decoherence and quantum to classical state transition \cite{zurek2003decoherence}. Earlier, general non-Gaussian state for macroscopic mechanical oscillator, have been demonstrated based on state transfer \cite{khalili2010preparing,akram2010single}, photon counting \cite{li2013enhancing}, photon subtraction \cite{paternostro2011engineering} and teleportation \cite{PhysRevA.83.013803} protocols. Other specific non-Gaussian states for mechanical oscillator are, fully developed Schr\"{o}dinger cat state \cite{tan2013deterministic,asjad2014reservoir,PhysRevA.91.013842,akram2013entangled} and Fock superposition state which has been developed by exploiting  the Kerr type nonlinearity \cite{PhysRevA.89.053829}.   

In this regard, putting a macroscopic mechanical oscillator into a non-Gaussian state, having one single quantum of energy, namely a single phonon Fock state (SPFS), is a sought after milestone. Such discrete energy eigen-state is important for the long distance quantum repeater setups as it shows very high fidelity and loss resilience \cite{sangouard2011quantum}. Moreover, realization of such state for macroscopic oscillator leads to the quantization of energy even at macroscopic scale. Thus, it directly manifests the implication of laws of quantum mechanic at macroscopic level. Recently, SPFS (quantum phonon state engineering) for mechanical oscillator has been proposed based on the parametric optomechanical interaction  and the photon counting \cite{galland2014heralded} (photon detection \cite{PhysRevLett.110.010504}) module. Also, such state has been realized experimentally by coupling the mechanical mode to a superconducting qubit \cite{o2010quantum}.
 
In this paper, we prepare single phonon Fock state (SPFS) of mechanical oscillator based on a coherent photon subtraction process from the field reflected by the mirror. Here we exploit the quantum mechanical correlation in negatively blue detuned regime i.e. $\Delta<0$, and the photon subtraction process is carried out on Stoke reflected field. Instantaneous measurement based on single photon subtraction process from the field, and the correlation set between field and mechanical modes, shapes the non-classicality of the mechanical mode. Identifying the nature of the obtained non-classicality for mechanical mode, we show that for appropriate set of parameters, the conditional mechanical state maps itself into SPFS. In order to check the effectiveness of result, we calculate the phonon number distribution and compare the obtained state to an ideal single quantum Fock state. Such a comparison has been quantified by a well accepted fidelity function \cite{tan2013deterministic,asjad2014reservoir}. We show that for a given instantaneous measurement on field, the  fidelity between obtained conditional mechanical state and the ideal single quantum Fock state approaches to unity. Moreover, phonon number distribution is sharply peaked around single quantum state and contribution made by other Fock states is highly negligible. Hence it manifests on demand engineering of SPFS of mechanical mode. 

This paper is organized as follows: In section \ref{sec:II}, we provide the model formulation. Section \ref{sec:III} contains dynamical equations of motion in linearized regime and a dynamical map for covariance matrix of two mode Gaussian state. In section \ref{sec:IV}, we provide analytical results for the conditional state and Wigner function of mechanical mode, post selected with the event of photon subtraction from the intra-cavity mode, and  we discuss the resultant conditional mechanical state. In section \ref{sec:V}, we explain our results for the obtained SPFS based on the Wigner function, fidelity and phonon statistics. Finally, in section \ref{S6}, we conclude our work.
 \section{The Model Formualtion}\label{sec:II}
\begin{figure}[t]\label{fig1}
\begin{center}
\includegraphics[width=0.4\textwidth]{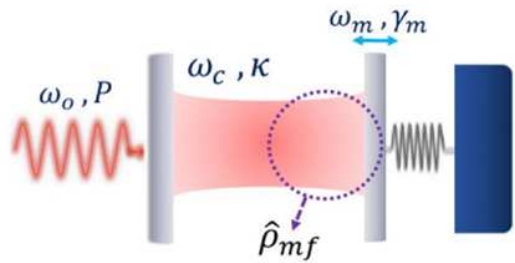}
 \caption{Paradigm model for optomechanical system. An optical mode of resonance frequency $\omega_{c}$ is confined between two mirrors which has decay rate $\kappa$. The mechanical mode associated with a movable mirror has frequency $\omega_{m}$ and coupled with its local environment with coupling rate $\gamma_m$. Cavity mode is excited with an input laser source of frequency $\omega_{o}$ and power $P$. The joint state of both modes is marked as $\hat{\rho}_{mf}$.}
 \label{fig:cavity}
\end{center}
\end{figure}

We consider a paradigm model of an optomechanical system, consisting of a cavity driven by an optical field mode of frequency $\omega_{o}$, with moving-end mirror as shown in Fig. \ref{fig:cavity}. Typical experimental configuration is the Farbry-P\'erot cavity with a movable mirror, but our scheme is well applicable to other systems such as photonic crystal nano-beam resonator \cite{chan2011laser}. We consider an intra-cavity field mode of frequency $\omega_{c}$, with decay rate $\kappa$ to its local environment, while the mechanical mode associated with the movable mirror is being modeled as harmonic oscillator of frequency $\omega_{m}$, which is coupled locally to its environment with rate $\gamma_{m}$ ~\cite{aspelmeyer2013cavity}. The optomechanical coupling is carried out with radiation pressure interaction ~\cite{law1995interaction}. In a frame rotating at laser frequency $\omega_{o}$, the Hamiltonian of the system reads ~\cite{vitali2007optomechanical},
\begin{equation}\label{eq:Hamiltonian}
 \hat{H}_{s}=\hbar\Delta_{0}\hat{a}^{\dagger}\hat{a}+\frac{\hbar\omega_{m}}{2}(\hat{q}^2+\hat{p}^2)-\hbar G_{0}\hat{q}\hat{a}^{\dagger}\hat{a}+i\hbar E(\hat{a}^{\dagger}+\hat{a}).
 \end{equation}
 Here, $\Delta_{0}=\omega_{c}-\omega_{0}$ is the detuning of the laser with respect to unperturbed cavity resonance frequency $\omega_{c}$. In Eq.~(\ref{eq:Hamiltonian}), the first term describes the free Hamiltonian of the single cavity field mode, where $\hat{a}(\hat{a}^{\dagger})$ is the corresponding annihilation(creation) operator. The second term accounts the Hamiltonian of the mechanical mirror, with dimensionless position and momentum coordinates for the mechanical oscillator are, $\hat{q}$ and $\hat{p}$ respectively. The third term illustrates the interaction Hamiltonian between the cavity field and moving mirror with single photon optomechanical coupling strength $G_{0}=G^{\prime}\sqrt{\frac{\hbar}{m\omega_{m}}}$. Here, $G^{\prime}=\omega_{c}/L$, $m$ is the mass of the oscillator and $L$ is the length of the unperturbed cavity. Finally the last term describes the laser interaction with the cavity mode with $E$ being the complex number connected to laser power $P$ by $|E|=\sqrt{\frac{2P\kappa}{\hbar\omega_{0}}}$. Based on the Hamiltonian in \eqref{eq:Hamiltonian}, we demonstrate the dynamical equations and characterize the statistical properties of the system by means of covarience matrix. 
\subsection{Equations of Motion}\label{sec:III}
In the context of open system dynamics under Heisenberg-Langevin formalism, the Hamiltonian in equation \eqref{eq:Hamiltonian} gives rise to a set of nonlinear differential equations of motion for the observables  $\hat{q}$, $\hat{p}$ and $\hat{a}$ of the system \cite{giovannetti2001phase}. For simplicity we omit the hat sign and write the equations of motion as,
 \begin{subequations}\label{eq:NLDE}
\begin{align}
\dot{q}(t)&=\omega_{m}{p},\label{eq:NLDE1}\\
\dot{p}(t)&=-\omega_{m}q+G_{0} a^{\dagger}a-\gamma_{m}p+\xi,\label{eq:NLDE2}\\
\dot{a}(t)&=-(\kappa+i\Delta_{0})a+iG_{0} q a+E+\sqrt{2\kappa} a_{in}.\label{eq:NLDE3}
\end{align}
\end{subequations}
Here, dot on each observable represents the derivative with respect to time, and $a_{in}$ is the input noise affecting the intracavity mode, characterized by only non-zero correlation viz. $\langle {a}_{in}(t){a}^{\dagger}_{in}(t^{\prime})\rangle=\delta(t-t^{\prime})$. In addition, $\xi$ is the zero mean Brownian stochastic force affecting the mechanical motion, which is due to phononic bath at temperature $T$. Inherently, this force is non-Morkovian in nature \cite{gardiner2004quantum}, however for large quality factor $Q=\omega_{m}/\gamma_{m}$, it is possible to restore its Morkovian nature such that it satisfies the correlation \cite{vitali2007optomechanical},
\begin{equation}\label{eq:brownian_noise}
\langle\xi(t)\xi(t^{\prime})+\xi(t^{\prime})\xi(t)\rangle\simeq\gamma_{m}(2\overline{n}+1)\delta(t-t^{\prime}).
\end{equation}
Here, $\overline{n}$ is the mean phonon number associated with the mechanical mode, and is expressed as $\overline{n}=(\exp[\hbar \omega_{m}/K_{B}T]-1)^{-1}$, where $K_{B}$ denotes the Boltzmann constant. Moreover, the operator $o=\lbrace q,p,a\rbrace$ in the form $o_{s}+\delta{o}$, and a large intracavity field amplitude make it possible to write c-number steady state parts as $q_{s}=G_{o}\vert a_{s}\vert ^{2}/\omega_{m}$, $p_{s}=0$, and $a_{s}=E/(\kappa+i\Delta)$. Furthermore, we write the following  set of linear coupled equations for fluctuation operator part $\delta{o}$ as~\cite{paternostro2006reconstructing},
\begin{equation}\label{eq:linerset}
\dot{\textbf{u}}(t)=\textbf{k}\textbf{u}(t)+\textbf{n}(t),
\end{equation}
where, $\textbf{u}=\left(\begin{array}{cccc}{\delta q}&{\delta p}&{\delta X}&{\delta Y}\end{array}\right)^{T}$ is the vector for quadratures of the system, $\textbf{n}=\left(\begin{array}{cccc}{o}&{\xi}&{\sqrt{2\kappa} X_{in}}&{\sqrt{2\kappa}Y_{in}}\end{array}\right)^{T}$ accounts for the Gaussian input noise vector, and $\textbf{k}$ is the system kernel matrix given by,
\begin{equation}\label{eq:kernal}
\textbf{k}=
\left(
\begin{array}{cccc}
{0}&{\omega_{m}}&{0}&{0}\\
{-\omega_{m}}&{-\gamma_{m}}&{G}&{0}\\
{0}&{0}&{-\kappa}&{\Delta}\\
{G}&{0}&{-\Delta}&{-\kappa}\\
\end{array}
\right).
\end{equation}
Here, $\delta X=(\delta a+\delta a^{\dagger})/\sqrt{2}$, $\delta Y=(\delta a-\delta a^{\dagger})/i\sqrt{2}$ are the field quadratures with their corresponding input fields $\delta X_{in}=(\delta a_{in}+\delta a_{in}^{\dagger})/\sqrt{2}$ and $\delta Y_{in}=(\delta a_{in}+\delta a_{in}^{\dagger})/i\sqrt{2}$.  
The parameter $G=\sqrt{2}\alpha_{s}G_{o}$ is the effective photon number coupling, while $\Delta=\Delta_{o}-G_{o}^2 \vert a_{s} \vert ^{2}/\omega_{m}$ is laser detuning with respect to effective cavity resonance frequency. We consider $ a_{s}$ to be real such that, $a_{s}=a_{s}^{\ast}\equiv\alpha_{s}=\vert E\vert/\sqrt{(\kappa^{2}+\Delta^{2})}$.
 The solution for equation \eqref{eq:linerset} is given by,
\begin{equation}\label{eq:linersetsol}
\mathbf{u}(t)=\textbf{M}(t)\mathbf{u}(0)+\int_{0}^{t}d\tau \textbf{M}(\tau)\textbf{n}(t-\tau), 
\end{equation}
where $\textbf{M}(t)=\exp(\mathbf{k}t)$, which tends to zero as $t\longrightarrow\infty$ for the stable output. We obtain the stability conditions by employing the Routh-Hurwitz criterion \cite{PhysRevA.35.5288}, which leads us to the following two nontrivial constraints on the system parameters given as,
\begin{align}
&c_{1}=\Delta  G^2 \omega _m \left(2 \kappa +\gamma _m\right){}^2+4 \kappa  \gamma _m \omega _m^2 \left(-\Delta ^2+\kappa ^2+\kappa  \gamma _m\right)+\nonumber\\
&2 \kappa  \left(\Delta ^2+\kappa ^2\right) \gamma _m \left(\Delta ^2+\kappa ^2+2 \kappa  \gamma _m+\gamma _m^2\right)+2 \kappa  \gamma _m \omega _m^4>0,\nonumber\\
&c_{2}=-\Delta  G^2 \omega _m+\Delta ^2 \omega _m^2+\kappa ^2 \omega _m^2>0.
\end{align}
For the rest of the analysis, we assume that the above two constraints are satisfied. In addition, the linear map given in equation \eqref{eq:linersetsol}, guarantees the output quadrature vector for joint state of field and mechanical modes to be Gaussian. Hence, we fully characterize the Gaussian state through its co-variance matrix $\textbf{v}(t)$, whose matrix elements are \cite{vitali2007optomechanical}, v$_{ij}(t)=\langle u_{i}(t)u_{j}(t)+u_{j}(t)u_{i}(t)\rangle/2 $, where $i,j={1,...,4}$. We write the co-variance matrix as,
\begin{equation}\label{eq:covarience}
\textbf{v}=\left(
\begin{array}{cccc}
{\textbf{m}}&{\textbf{c}}\\
{\textbf{c}^T}&{\textbf{f}}\\
\end{array}
\right)\equiv\left(
\begin{array}{cccc}
{m_{11}}&{m_{12}}&{c_{11}}&{c_{12}}\\
{m_{21}}&{m_{22}}&{c_{21}}&{c_{22}}\\
{c_{11}}&{c_{21}}&{f_{11}}&{f_{12}}\\
{c_{12}}&{c_{22}}&{f_{21}}&{f_{22}}\\
\end{array}
\right)
\end{equation}
where, the sub-matrices $\textbf{m}$ and $\textbf{f}$, hold the local properties of mechanical mode and field mode respectively, while any type of correlation between these two subsystem is encompassed by the matrix $\textbf{c}$.
At this point, we write equation of motion for the covariance matrix which reads \cite{rogers2014hybrid,ferreira2009quantum},
\begin{equation}\label{eq:eomcovar}
\dot{\textbf{v}}(t)=\textbf{k}\textbf{v}(t)+\textbf{v}(t)\textbf{k}^{T}+\textbf{D}.
\end{equation}
Here it is natural to select the initial joint state of mechanical and field modes to be in thermal and vacuum states respectively. This is given by initial joint covariance matrix $\textbf{v}(0)=(\overline{n}+\frac{1}{2})\mathbb{I}_{2}\bigoplus\frac{1}{2}\mathbb{I}_{2}$ \cite{rogers2012entanglement}, where $\mathbb{I}_{2}$ is a $(2\times2)$ identity matrix. Moreover, $\overline{n}$ is the initial occupancy of phonon for the mechanical mode set by the environmental temperature and mechanical frequency. In the following section, we construct the joint state of field and mechanical modes based on the joint covariance matrix. Furthermore, we recast the state of mechanical mode upon subtracting a single photon from the cavity mode. 
\subsection{Reconstruction of the Wigner Function for Mechnical Mode} \label{sec:IV}
As discussed earlier, the main goal of the scheme is to subtract a single photon from the field reflected by the mirror at a given instant of interaction time.  In this way, we recast the state of mechanical mode by making a measurement on reflected field which is based on photon subtraction process. In order to quantify the conditional state of mechanical mode after subtracting a photon from the cavity mode, we formulate the time resolved conditional Wigner function of the mechanical mode. 

Enlightening that density operator  $\hat{\rho}$ corresponding to physical state is bounded, i.e. its Hilbert Schmidt norm is finite \cite{cahill1969ordered, kenfack2004negativity}, which makes possible to expand it in the basis $\lbrace\hat{D}^{\dagger}(\lambda)\forall\lambda\in\mathbb{C}, \lambda=\lambda_{r}+i\lambda_{i} \rbrace$. Hence we write,
\begin{equation}\label{singledensity}
\hat{\rho}\equiv \frac{1}{\pi}\int d^{2}\lambda C(\lambda,\lambda^{\ast})\hat{D}^{\dagger}(\lambda).
\end{equation}
Here $\hat{D}^{\dagger}(\lambda)$ is the conjugate transpose of coherent displacement operator $\hat{D}(\lambda)\equiv exp[\lambda \hat{a}^{\dagger}+\lambda^{\ast}\hat{a}]$ and $d^{2}\lambda=d\lambda_{r} d\lambda_{i}$ is the differential for two dimensional integral corresponding to real part $\lambda_{r}$ and imaginary part $\lambda_{i}$ of the complex variable $\lambda$. Moreover, we define Weyal characteristic function $C(\lambda,\lambda^{\ast})$ as an expectation value of displacement operator, which is given by $C(\lambda,\lambda^{\ast})\equiv tr[\hat{\rho} \hat{D}(\lambda)]$. Here, $tr[.]$ represents the trace operation. With the above definitions, we write two mode version of equation \eqref{singledensity} for the joint mirror-field state $\hat{\rho}_{mf}$ \cite{kim2005nonclassicality} as,
\begin{equation}
\hat{\rho}_{mf}=\frac{1}{\pi^{2}}\int d^{2}\lambda d^{2}\eta C_{mf}(\lambda,\eta)\hat{D}^{\dagger}_{m}(\eta)\otimes\hat{D}_{f}^{\dagger}(\lambda).
\end{equation}
Here, the two dimensional deferential is given by $d^{2}a=da_{r}da_{i}$, where $a=\lambda,\eta$. In addition,  $\hat{D}^{\dagger}_{m}(\eta)$ and $\hat{D}_{f}^{\dagger}(\lambda)$ are the displacement operators, corresponding to mechanical mode and field mode respectively, and $\textbf{x}\equiv[\lambda_{r},\lambda_{i},\eta_{r},\eta_{i}]$ is the vector of real variables for the four dimensional phase space of the system. At any instant of time, the joint state of these two modes is Gaussian which is characterized by the time dependent co-variance matrix given in Eq. \eqref{eq:covarience}. Hence, using the co-variance matrix, we write time dependent Weyal characteristic function as $C_{mf}(\lambda,\eta,t)\equiv exp[-\frac{1}{2}\textbf{x}\textbf{v}(t)\textbf{x}^{T}]$. 

To get the instantaneous conditional state of the mechanical mode upon photon subtraction, we linearly transform field displacement operator by annihilation operator, which results in the a photon subtracted field. Furthermore, we take partial trace on the field mode in order to eliminate the field variables. Such a process is equivalent to make a measurement on field mode for the observation of one field quanta \cite{paternostro2011engineering,kim2005nonclassicality, li2013enhancing}. In this way, our conditional density operator for mechanical mode becomes, 
\begin{equation}\label{conditionalsta}
\hat{\rho}_{m}=\frac{\aleph}{\pi^{2}}\int d^{2}\lambda d^{2}\eta C_{mf}(\lambda,\eta)\hat{D}^{\dagger}_{m}(\eta)tr[\hat{a}\hat{D}_{f}^{\dagger}(\lambda)\hat{a}^{\dagger}].
\end{equation}
Here, $\aleph$ is the normalization constant, and the quantity $tr[\hat{a}\hat{D}_{f}^{\dagger}(\eta)\hat{a}^{\dagger}]$ reads,
\begin{align}\label{tracecal}
tr[\hat{a}\hat{D}_{f}^{\dagger}(\eta)\hat{a}^{\dagger}]=\frac{1}{\pi}\int& d^{2}\alpha[\vert\alpha\vert^{2}-\vert\lambda\vert^{2}+\lambda^{\ast}\alpha-\alpha^{\ast}\lambda+1]\nonumber\\
&exp[-\frac{1}{2}\vert\lambda\vert^{2}+\lambda^{\ast}\alpha-\alpha^{\ast}\lambda].
\end{align}
Using the above expression in Eq.~\eqref{conditionalsta} we are left with,
\begin{equation}\label{conditionalstate1}
\hat{\rho}_{m}=\frac{\aleph}{\pi^{2}}\int d^{2}\eta d^{2}\alpha\hat{D}^{\dagger}_{m}(\eta)F_{or}[g(\lambda,\alpha,\eta)].
\end{equation}
Here, $F_{or}[\cdot]$ is the complex Fourier transform of function $g=g(\lambda,\alpha,\eta)$. The complex Fourier transform \cite{cahill1969ordered} of any function $h(\xi)$, results in the function $f(\beta)$ such that $f(\beta)=F_{or}[h(\xi)]$. Explicitly, it is given by $f(\beta)\equiv\pi^{-2}\int d^{2}\xi h(\xi)\exp(\beta\xi^{\ast}-\beta^{\ast}\xi)$. With this definition, the function $g(\lambda,\alpha,\eta)$ in equation \eqref{conditionalstate1}, takes the form,
\begin{equation}\label{fourierfun}
g(\lambda,\alpha,\eta)= C_{mf}(\lambda,\eta)[\vert\alpha\vert^{2} - \vert\lambda\vert^{2} + \lambda^{\ast}\alpha-\alpha^{\ast}\lambda+1]e^{-\frac{1}{2}\vert\lambda\vert^{2}}.
\end{equation}
The Fourier transform over variable $\lambda$ in Eq.~(\ref{conditionalstate1}), followed by the integration over variable $\alpha$, results in the conditional density operator for mechanical mode which is given by,
\begin{equation}\label{conditionalstate}
\hat{\rho}_{m}=\frac{\aleph}{4\pi}\int d^{2}\eta \hat{D}^{\dagger}_{m}(\eta)\exp[g_{1}(\eta)][g_{2}(\eta)],
\end{equation}
where,
\begin{equation}
g_{1}(\eta)=-\frac{1}{2} m_{22} \eta _i^2-\frac{1}{2} \eta _r \left(m_{12} \eta _i+m_{21} \eta _i+m_{11} \eta _r\right),
\end{equation}
\begin{align}
&g_{2}(\eta)=-c_{21}^2 \eta _i^2-c_{22}^2 \eta _i^2-2 c_{11} c_{21} \eta _i \eta _r-2 c_{12} c_{22} \eta _i \eta _r\nonumber\\
&-c_{11}^2 \eta _r^2-c_{12}^2 \eta _r^2+f_{11}+f_{22}-2.
\end{align} 
Any type of effect due to photon subtraction on mechanical mode is then carried out by two functions, namely $g_{1}(\eta)$ and $g_{2}(\eta)$. We further get the conditional mechanical characteristic function \cite{cahill1969ordered}, which has definition $C_{m}(\gamma)\equiv tr[\hat{\rho}_{m}\hat{D}(\gamma)]$. In this way, our conditional characteristic function for mechanical mode becomes,
\begin{equation}\label{conchrac}
C_{m}(\gamma)=\frac{\aleph}{4}\exp[g_{1}(\gamma)][g_{2}(\gamma)].
\end{equation}
We further get the conditional Wigner function, $W(\delta_{r},\delta_{i})$, of the mechanical mode by taking complex Fourier transform of the characteristic function, given by  $W(\delta)=\pi^{-2}\int d^{2}\gamma C_{m}(\gamma)\exp(\gamma^{\ast}\delta-\gamma\delta^{\ast})$. On this transformation, we write the explicit form of conditional Wigner function for mechanical mode as,
\begin{align}\label{wigner}
&W(\delta_{r},\delta_{i})=A_{1}\exp [A_{2}] [B_{1}-\frac{B_{2} B_{3}}{C^{2}}-\frac{B_{4}B_{5}}{C}+B_{6}].
\end{align}
Here, $A_{1}$ contains the normalization constant expressed as,
\begin{equation}\label{normalization}
A_{1}=\frac{\aleph}{\pi }\left(\frac{1}{m_{11}}\right){}^{5/2} \sqrt{\frac{m_{11}}{4 m_{11} m_{22}-\left(m_{12}+m_{21}\right){}^2}},
\end{equation}
and exponent $A_{2}$ takes the form,
\begin{equation}
A_{2}=\frac{8 \left(m_{22} \delta _i^2+\left(m_{12}+m_{21}\right) \delta _i \delta _r+m_{11} \delta _r^2\right)}{\left(m_{12}+m_{21}\right){}^2-4 m_{11} m_{22}}.
\end{equation}
Moreover, in Eq.~(\ref{wigner}), we have a set of polynomials over two dimensional phase space variables $\delta_{r}$ and $\delta_{i}$ given by, 
\begin{equation}
B_{1}=-\left(c_{11}^2+c_{12}^2\right) \left(m_{11}-4 \delta _i^2\right),
\end{equation}
\begin{widetext}
\begin{equation}
B_{2}=4 \left(c_{21}^2+c_{22}^2\right) m_{11}^2-4 \left(c_{11} c_{21}+c_{12} c_{22}\right) \left(m_{12}+m_{21}\right) m_{11}+\left(c_{11}^2+c_{12}^2\right) \left(m_{12}+m_{21}\right){}^2, 
\end{equation}
\begin{equation}
B_{3}=-4 \left(m_{12}+m_{21}\right){}^2 \delta _i^2-\left(m_{12}+m_{21}\right) m_{11} \left(16 \delta _i \delta _r+m_{12}+m_{21}\right)+4 m_{11}^2 \left(m_{22}-4 \delta _r^2\right),
\end{equation}
\begin{equation}
B_{4}=8 \left(c_{11}^2 \left(m_{12}+m_{21}\right)-2 c_{21} c_{11} m_{11}+c_{12} \left(c_{12} \left(m_{12}+m_{21}\right)-2 c_{22} m_{11}\right)\right),
\end{equation}
\end{widetext}
\begin{equation}
B_{5}=\left(m_{12}+m_{21}\right) \delta _i^2+2 m_{11} \delta _i \delta _r,
\end{equation}
\begin{equation}
B_{6}=\left(f_{11}+f_{22}-2\right) m_{11}^2,
\end{equation}
\begin{equation}\label{eqlastpoly}
C=\left(m_{12}+m_{21}\right){}^2-4 m_{11} m_{22}.
\end{equation} 
Finally, the normalization constant $\aleph$  in \eqref{normalization} reads,
\begin{equation}
\aleph=\frac{4}{f_{11}+f_{22}-2}.
\end{equation}
In expression \eqref{wigner}, the obtained conditional Wigner function clearly indicates that it is no more Gaussian in its phase space. Depending on the polynomials, given by equations \eqref{normalization} to \eqref{eqlastpoly}, this function can take negative values which is the signature of non-classicality \cite{hudson1974wigner}. Furthermore, from this expression, we note that it depends on time evolved values of the co-variance matrix elements. Since measurement made on field mode at a dynamical instant of time, sets the instantaneous values for the covariance matrix. Therefore, it is obvious to get a particular conditional state of mechanical mode at that instant of time. We show in the following analysis that, for given set of parameters and at a specific dynamical instant of time, the conditional mechanical state obtained by this  Wigner function maps itself into a SPFS.
\section{Results and Discussions}\label{sec:V}
Following the linearized regime of optomechanical set-up, in our numerical results, we preserve the experimentally realisable week coupling condition \cite{galland2014heralded}, which is followed by the argument that optomechanical coupling must be less than local cavity decay rate i.e. $G<\kappa$. In addition, we assume that the system operates in the resolved side band regime such that $\kappa<\omega_{m}$ holds for our system. Our scheme also encompasses the possibility of on-demand photon subtraction at well defined instant by passing a two-level atom with appropriate conditions under Jaynes Cummings Paul type interaction \cite{kim2008recent,rogers2014hybrid,RevModPhys.85.1083,ANDP:ANDP201400107}.


In our numerical results, we use experimentally accessible parameteric values \cite{chan2011laser, chan2012optimized,galland2014heralded}: $L=1$ mm, $\lambda=1064$ nm, $\omega_{m}/2\pi=1$ GHz, $P=5$ mW, $m=5$ ng, $F=10^4$, $T=1$ mK, $\omega_{c}=\omega_{o}$, and $\gamma_{m}/2\pi=100$ Hz. In this set, main controllable parameter is the mechanical frequency $\omega_{m}$, which is on the order of GHz range, quite necessary for a high quality mechanical mode. Earlier a GHz frequency range for mechanical mode has been utilized to create single phonon state \cite{galland2014heralded}, and reported experimentally in photonic crystal nano-beam resonator \cite{chan2012optimized}. We work in the blue detuned regime \cite{genes2008robust} i.e. $\Delta<0$ (in the units of mechanical frequency), where laser frequency has larger value than the effective cavity resonance frequency. Such operational regime causes extra heating of mechanical mode by enhancing its quanta \cite{hammerer2014nonclassical} in general, and system moves toward instability. Here we use the set of parameters which prevent extra heating of the mechanical mode and system remains stable. With these parameters, $G$ takes the value $51.847$ KHz, whereas the optical line-width viz. $\kappa=\frac{\pi c}{FL}$, takes the value $94$ MHz. Hence, it places the setup into the weak coupling condition with $G/\kappa\simeq0.00055$. In addition, provided that stoke side band mode is fully resonant with cavity resonance, the system operates in the resolved sideband regime by fulfilling the condition $\kappa<\omega_{m}$.

\begin{figure}[t]
\begin{center}
\includegraphics[width=0.4\textwidth]{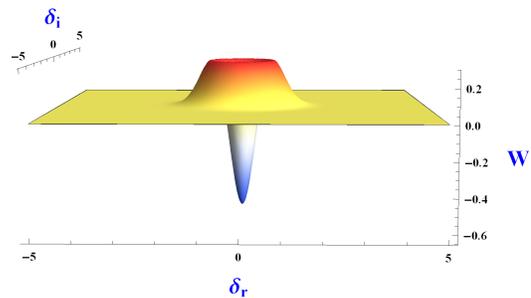}
 \caption{Wigner representation of obtained conditional mechanical state for parameters: $L=1$ mm, $\lambda=1064$ nm, $\omega_{m}/2\pi=1$ GHz, $P=5$ mW, $m=5$ ng, $F=10^4$, $T=1$ mK, $\omega_{c}=\omega_{o}$, and $\gamma_{m}/2\pi=100$ Hz. We take $\Delta/\omega_{m}=-1$ and measurement is made at interaction time $9$ $\mu$s.}
 \label{fig:wignercond}
\end{center}
\end{figure}

\begin{figure}[t]
\begin{center}
\includegraphics[width=0.4\textwidth]{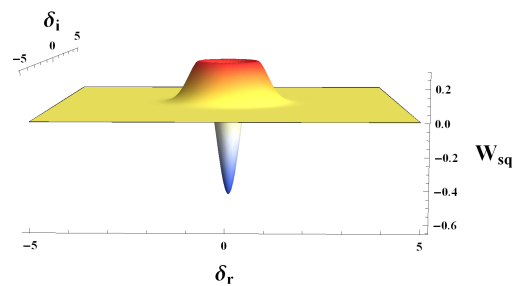}
 \caption{Wigner representation for an ideal single quantum Fock state, given in expression \eqref{idealfock}.}
  \label{fig:wigneridel}
\end{center}
\end{figure}

With these realizable experimental parametric values, we now discuss our main numerical results. In fact at a selective time instant of $9$ $\mu$s for the measurement, we obtain the Wigner function of resultant conditional mechanical state according to expression (\ref{wigner}), which is shown in Fig.  \ref{fig:wignercond}. Remarkably, we observe a high rotational symmetry around the origin of the phase space. The rotational symmetry around the origin is an essential characteristic of Fock states of bosonic modes in their Wigner representation \cite{schleich2011quantum}. Furthermore, it shows a high resemblance with an ideal single quantum Fock state.  For comparison, we present the Wigner representation $W_{sq}$ for an ideal single quantum Fock state as shown in Fig. \ref{fig:wigneridel}, given by the expression \cite{agarwal2012quantum,biswas2007nonclassicality},
\begin{equation}\label{idealfock}
W_{sq}(\delta)=\left\{\frac{2 e^{-2 \delta _i^2-2 \delta _r^2} \left\{4 \delta _i^2+4 \delta _r^2-1\right\}}{\pi }\right\}.
\end{equation}

 \begin{figure}[t]
\begin{center}
\includegraphics[width=0.4\textwidth]{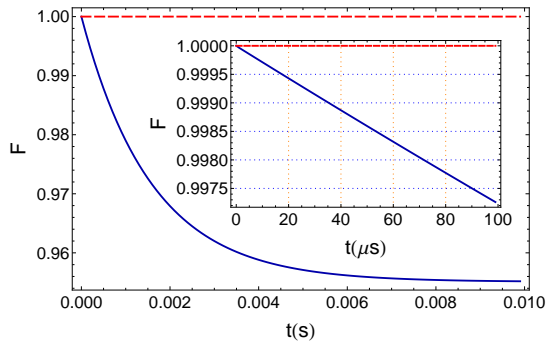}
 \caption{Conditional fidelity for different measurement time instants. Here, $F\rightarrow1$ in transient regime and deviates from unity for measurement taken in steady state times. The inset shows the conditional fidelity in transient regime measurements with time given in  microsecond $(\mu s)$ scale. All the parameters are the same as in Fig. \ref{fig:wignercond}.}
 \label{fig:fedility1}
\end{center}
\end{figure}
\begin{figure}[t]
\begin{center}
\includegraphics[width=0.3\textwidth]{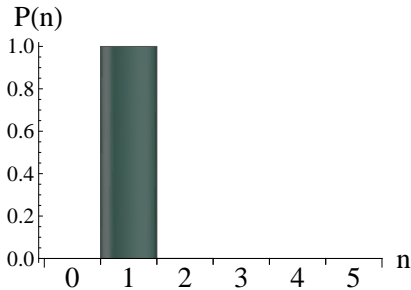}
 \caption{Phonon number distribution for the obtained state is shown. All the parameters are the same as in Fig. \ref{fig:wignercond}.}
\label{fig:phostat}
\end{center}
\end{figure}  

In this way, we see a clear closeness between two states in Fig. \ref{fig:wignercond} and Fig. \ref{fig:wigneridel}.
For the quantitative measure of the closeness of our prepared state with that of single quantum Fock state, we use a widely accepted measure which is $F$, the fidelity  function \cite{kim2008recent}. This function is simply the overlap of two states and can be defined if at least one of the states is pure. When two states overlap completely, fidelity $F$ takes the unit value. Here, we define conditional fidelity as the overlap of obtained conditional mechanical state to an ideal single quantum Fock state. In terms of Wigner functions, this is given by,
\begin{equation}
F=\pi\int d^{2}\delta W(\delta)W_{sq}(\delta).
\end{equation}

Using equations \eqref{wigner} and \eqref{idealfock}, we numerically compute the conditional fidelity versus the  measurement time which is shown in Fig. \ref{fig:fedility1}. We observe that fidelity approaches to unity for measurement taken in transient regime. It optimizes to a value $F\sim0.99974$, for measurement time $9$ $\mu$s, at which snapshot of conditional Wigner function in Fig. \ref{fig:wignercond} has been captured. For an intuitive quantification of conditional fidelity and measurement time instant in the transient regime, we show the numerical data in table \ref{table:nonlin}. We see that, for arbitrarily chosen measurement time instants in micro second range, the fidelity is optimized to $F>0.999$. Moreover, as time increases, the fidelity starts to decrease at fourth decimal place. It decreases from 0.999 value at t$\simeq$35 $\mu$s (see inset of Fig. \ref{fig:fedility1}).  Moreover, for an experimental convenience, a delayed measurement time is always desired. Hence it is appropriate to select the microsecond range for the measurement \cite{galland2014heralded} such that fidelity has more closer value to unity as well as the setup remains experimentally feasible for a delayed measurement. These time regimes then sets the criteria of $F>0.999$, for obtaining SPFS on the basis of fidelity.

In order to further reinforce the results, we calculate phonon statistics of the resultant mechanical state,  viz. $P(n)=\langle n\vert\hat{\rho}_{m} \vert n \rangle$  \cite{kim2008recent} and illustrated in the Fig. \ref{fig:phostat}. Here the distribution corresponding first five eigen states is shown for the same set of parameters. We see that distribution is sharply peaked for $n=1$ state, and contribution made by other Fock states is almost absent. This ensures the occurrence of single phonon Fock state of mechanical mode.

\begin{table}[t]
\caption{Conditional fidelity verses measurement time in transient regime for same parameters given in Fig. \ref{fig:wignercond}. }  
\centering          
\begin{tabular}{c c c c}    
\hline\hline\\[0.02ex] 
$t(\mu s)$ & $F$ & $t(\mu s)$ & F \\[0.2ex]  
\hline    

1.0001 & 0.999968 & 25.0001 & 0.999289  \\      
5.0001 & 0.999854 & 30.0001 &0.999149 \\
10.0001 & 0.999712 & 35.0001 & 0.999009  \\
15.0001 & 0.999571 & 40.0001 &0.99887 \\
20.0001 &0.99943& 45.0001 & 0.998731\\[2ex]  
\hline          
\end{tabular}
\label{table:nonlin}    
\end{table} 

\begin{figure}[t]
\begin{center}
\includegraphics[width=0.4\textwidth]{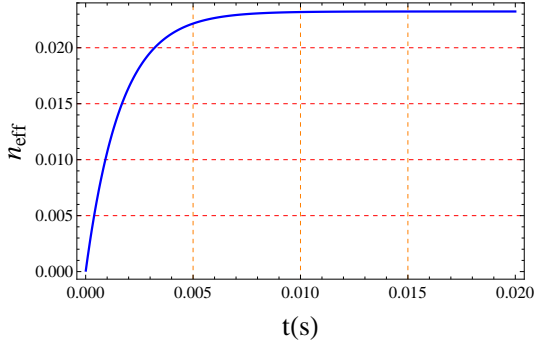}
 \caption{We show the effective phonon number profile as a function of interaction time. The convergence of phonon number up to steady state time has been shown. All the parameters are the same as in Fig. \ref{fig:wignercond}.}
 \label{fig:effphonon}
\end{center}
\end{figure}

Emergence of SPFS of mechanical mode via photon subtraction can be understood as follows: In fact, for negative blue detuned regime $\Delta=-\omega_{m}$, the optomechanical system follow the two mode optomechanical parametric interaction \cite{paternostro2008mechanism,genes2008robust} quite appropriate and manifests the optomechanical entanglement. Moreover, in this regime, entanglement evaluated through logarithmic negativity $E_{N}$ is bounded by $E_{N}\leq ln2$ and any thermally excited quanta, viz.  $n_{th}>1$, destroy the existence of entanglement because of instability \cite{hammerer2014nonclassical}. Here, low temperature limit, a high quality mechanical mode and low power coupled to the far off resonant mode, compensate this condition of instability. For clarity, we show the effective phonon number profile $n_{eff}=(m_{11}+m_{22}-1)/2$ of mechanical mode as a function of interaction time in Fig. \ref{fig:effphonon}, for the same set of parameters. Here we can see the convergence of effective phonon number to a finite number i.e. $n_{eff}<1$, even upto a steady state interaction time. In fact, the low temperature limit of bath upto $1$ mK and a high quality mechanical mode, facilitates that extra heating is avoided by mechanical mode. We add here that, apart from the argument of lower bath temperature, a small initial occupancy of mechanical phonon number can be achieved first by pre-cooling the mechanical mode by a red detuned field \cite{galland2014heralded}.
\begin{figure}[t]
\begin{center}
\includegraphics[width=0.4\textwidth]{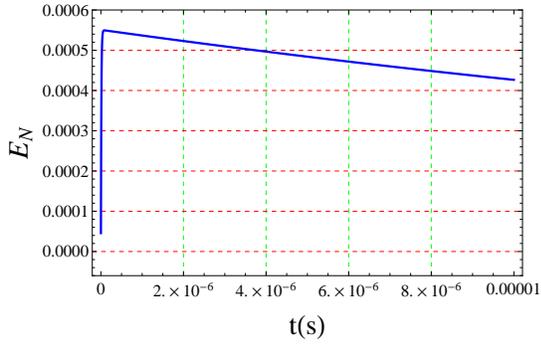}
 \caption{Dynamical entanglement through logarithmic negativity function is shown. Existence of optomechanical entanglement is verified in the transient regime as $E_{N}>0$. All the parameters are the same as in Fig. \ref{fig:wignercond}.}
 \label{fig:entanglement}
\end{center}
\end{figure}

\begin{figure}[t]
\begin{center}
\includegraphics[width=0.7\linewidth]{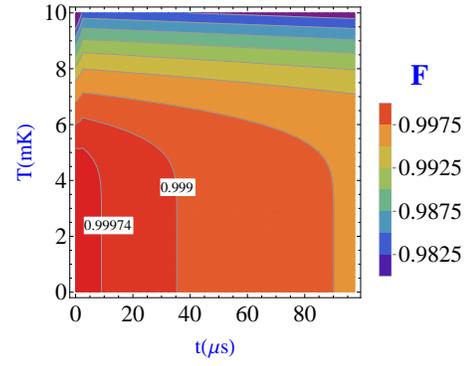}
\caption{Conditional fidelity against temperature and measurement time is shown. Note that, fidelity decreases as we take measurement with large temperature values. All the parameters are the same as in Fig. \ref{fig:wignercond}.}
\label{fig:fedilitya}
\end{center}
\end{figure} 

\begin{figure}[t]
\begin{center}  
\includegraphics[width=0.4\textwidth]{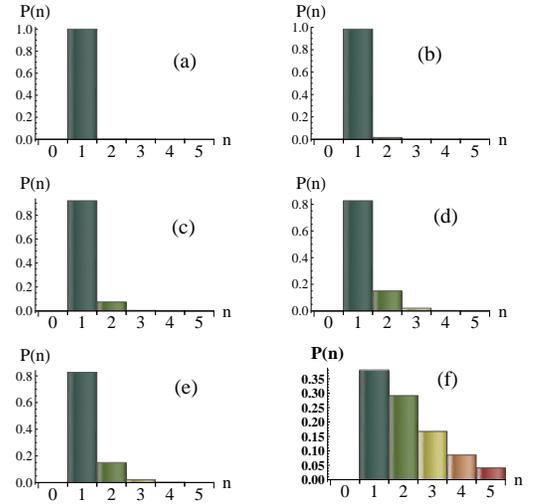}
\caption{Variation of phonon number distribution for the different values of temperature is shown for: (a) $5$ mK (b) $10$ mK (c) $15$ mK (d) $20$ mK (e) $25$ mK, and (f) $50$ mK. All the other parameters are the same as in Fig. \ref{fig:wignercond}.}
 \label{fig:phostattemp}
\end{center}
\end{figure} 
    
 With these parameters, namely, a high quality mechanical mode and low temperature limit used here, carries a unitary approach for optomechanical interaction for its time evolution \cite{paternostro2008mechanism}. In this way, the dynamics defines a two mode squeezing of optical and mechanical mode such that an entangled  photon-phonon  pair generates at an instantaneous value of interaction. We ensure the existence of dynamical entanglement $E_{N}$ by drawing logarithmic negativity function  \cite{rogers2014hybrid} in Fig. \ref{fig:entanglement} for the same set of parameters as discussed above. Here, existence of entanglement in transient regime of optomechanical interaction is verified by identifying the positive values for logarithmic negativity function i.e. $E_{N}>0$ and takes the value just above 0.0004 for 9 $\mu$s. Hence, it manifests a clear signature of two mode squeezing of the optical and mechanical modes.

While system achieves a two mode squeezed state at a dynamical instant of interaction, we ensure a measurement like photon subtraction on field, which is effectively equivalent to the observation of one filed quanta \cite{li2013enhancing,paternostro2011engineering}. This recasts the state of mechanical mode to a SPFS. Similar schemes has also been used for the generation of single and two photon Fock state of field by conditional measurement on the photons pair born in the parametric down conversion process for optical modes \cite{lvovsky2001quantum,zavatta2004tomographic,ourjoumtsev2006quantum}.

In this way, similarity seen between two states, namely the obtained state and the standard single quantum Fock state,  is followed by point that the conditional Wigner function of obtain state depends on the time evolved version of co-variance matrix. As the co-variance matrix achieves the two mode squeezing form, the expression of conditional Wigner function in \eqref{wigner} reduces to the expression \eqref{idealfock} of an ideal single quantum Fock state and fidelity approaches to unit value. Furthermore, as shown in Fig. \ref{fig:fedility1}, fidelity decreases largely when the system enters into the steady state time regimes. This is because of the fact that, as optomechanical interaction time evolves, the thermal bath associated with mechanical mode starts to play its role. The unitary evolution for two mode squeezing interaction becomes lesser effective and system promotes an open system dynamics of two bosonic modes. The conditional mechanical state based on the measurement made in this regime deviates from the single phonon state. This causes a large deviation of conditional Fidelity from unit value in steady state times (see for instance, Ref. \cite{asjad2014reservoir}). 
  
 In order to see the effect of initial thermal occupancy, we study the effect of temperature on the obtained state as shown in Fig. \ref{fig:fedilitya}. Here, we present the conditional fidelity against temperature and measurement time.  We specifically show the two cotours of $F=0.99974$ and $F=0.999$ values. Former reflects the values of time and temprature at which the Wigner function of the conditional mechnical state has been shown in Fig. \ref{fig:wignercond}, while later specifics the values of these parameters for $F\geq0.999$ criteria. 
 
 In general, we observe that on increasing the bath temperature, the fidelity decreases. This reflects that the obtained state deviates from the SPFS. This is because with larger bath temperature, system leaves the unitary evolution and the resultant conditional state moves towards a non-classical state whose statistics is significantly different from a SPFS. This can be seen form Fig. \ref{fig:phostattemp}(a-f), where we show the phonon number distribution for different values of bath temperature. We observe that, as the bath temperature increases, the distribution of phonon number deviates from $n=1$ state. For $T>10$ mK and upto $50$ mK, the mechanical state emerges with arbitrary population. These results suggest that our optimum range for achieving SPFS would be a transient regime of optomechanical interaction with small value of temperature i.e. a pre-cooled mechanical mode.

In the context of above discussions, our studies provide a unique platform to prepare a single phonon fock state of macroscopic mechanical mode via photon subtraction process. This work has potential applications in long distance quantum communication networks, quantum repeaters, quantum memories and quantum information processing based on non-classical mechanical states \cite{sangouard2011quantum,galland2014heralded,stannigel2012optomechanical,aspelmeyer2013cavity}.   
 
\section{Conclusions}\label{S6}
We have proposed a scheme for the preparation of single phonon Fock state of massive mechanical mode based on photon subtraction process. We provide full analytical formulation of Wigner function, by keeping the off diagonal terms in local covariance matrix, which is handy for identifying the measurement time for the formation of SPFS in transient regime of optomechanical interaction. The confirmation of SPFS has been carried out through conditional fidelity and phonon number distribution. We have shown that, for a given set of parameters and at fix instant of interaction time, the obtained state overlaps to an ideal single quantum state. Our present scheme carries the SPFS confirmation based on fidelity and phonon distribution which encompass co-variance matrix elements and Wigner function. This makes an experimental convenience as both are  reconstructable via high-precision all-optical procedures \cite{mari2011directly, laurat2005entanglement}, and mechanical state tomography \cite{vanner2013cooling}. In this way, we present a road map to engineer the SPFS of massive mechanical mode and its confirmation based on continuous variable quantum optics tools.

\begin{acknowledgments}
We submit our thanks to M. Paternostro for valuable suggestions in our manuscript. We acknowledge Higher Education Commission, Pakistan and Quaid-i-Azam University for financial support through Grants No. \#HEC/20-1374, and No. QAU-URF2014. 
\end{acknowledgments}

\bibliographystyle{apsrev}
\bibliography{References}
\end{document}